\documentclass[superscriptaddress,aps,pra,twocolumn]{revtex4}
\usepackage{bm}
\usepackage{ulem}
\usepackage{epsfig}
\usepackage{graphicx}
\usepackage{amssymb,amsmath,amsbsy,amsgen,amsfonts}    
\usepackage{dcolumn}
\usepackage{amsthm}
\usepackage{mathrsfs}
\usepackage{latexsym}
\usepackage{array}
\usepackage{color}
\usepackage{amstext}
\allowdisplaybreaks[1]
\usepackage{txfonts}
\usepackage{pifont}
\usepackage{booktabs,multirow}
\usepackage{mathtools}
\usepackage{hhline}

\usepackage{epstopdf} 

\newcommand{\bigbraket}[1]{\left\langle#1\right\rangle}
\newcommand{\braket}[1]{\langle#1\rangle}

\newcommand{\abs}[1]{|#1|}

\newcommand{\bra}[1]{\left\langle{#1}\right\vert}
\newcommand{\ket}[1]{\left\vert{#1}\right\rangle}

\DeclareSymbolFont{symbols}{OMS}{cmsy}{m}{n}

\begin{document}
\title{Sub-Poisson-Binomial Light}

\author{Changhyoup Lee}
\email{changdolli@gmail.com}
\affiliation{Institute of Theoretical Solid State Physics, Karlsruhe Institute of Technology, 76131 Karlsruhe, Germany}

\author{Simone Ferrari}
\author{Wolfram H. P. Pernice}
\affiliation{Institute of Physics, University of Muenster, 48149 M\"unster, Germany}

\author{Carsten Rockstuhl}
\email{carsten.rockstuhl@kit.edu}
\affiliation{Institute of Theoretical Solid State Physics, Karlsruhe Institute of Technology, 76131 Karlsruhe, Germany}
\affiliation{Institute of Nanotechnology, Karlsruhe Institute of Technology, 76021 Karlsruhe, Germany}

\date{\today}

\begin{abstract}
We introduce the general parameter $Q_{\rm PB}$ that provides an experimentally accessible nonclassicality measure for light. The parameter is quantified by the click statistics obtained from on-off detectors in a general multiplexing detection setup. Sub-Poisson-binomial statistics, observed by $Q_{\rm PB}<0$, indicates that a given state of light is nonclassical. 
Our new parameter replaces the binomial parameter $Q_{\rm B}$ for more general cases, where any unbalance among the multiplexed modes is allowed, thus enabling the use of arbitrary multiplexing schemes.
The significance of the parameter $Q_{\rm PB}$ is theoretically examined in a measurement setup that only consists of a ring resonator and a single on-off detector. The proposed setup exploits minimal experimental resources and is geared towards a fully integrated quantum nanophotonic circuit. The results show that nonclassical features remain noticeable even in the presence of significant losses, rendering the new nonclassicality test more practical and flexible to be used in various nanophotonic platforms.
\end{abstract}


\maketitle

\section{introduction}
Experimentally measurable tests for the nonclassicality of light are important not just for fundamental studies but also for many quantum information applications \cite{Bouwmeester97,Gisin07,Ladd10}. Various schemes have been proposed to identify light fields that require a quantum mechanical description \cite{Dodonov00, Richter02, Kenfack04, Asboth05}. One prominent nonclassical feature of single-mode electromagnetic field is the sub-Poissonian photon number statistics, which can be distinguished by measuring the Mandel parameter $Q_{\rm M}$ \cite{Mandel79}, defined as
\begin{equation}
Q_{\rm M}=\frac{\langle (\Delta n)^{2} \rangle}{\langle n \rangle}-1.
\label{QM}
\end{equation}
Here, $\langle n\rangle$ and $\langle (\Delta n)^{2} \rangle$ represent the classical mean value and variance of the photon number statistics. It has been shown that states showing a sub-Poisson statistics, identified by $Q_{\rm M}<0$, are nonclassical. This provides a simple nonclassicality test. However, measuring $Q_{\rm M}$ crucially relies on the availability of accurate photon-number-resolving detectors. Recently, such detectors became available \cite{Gerrits10, Namekata10, Gerrits11, Brida12} but more accessible and versatile schemes are still required in many cases where complex quantum devices are considered.

To circumvent the necessity of having detectors available with photon-number-resolving capability, a multiplexing strategy based on the principle of divide-and-conquer has been suggested to measure the nonclassical photon-number statistics using on-off detectors \cite{Banaszek03, Rehacek03, Achilles03, Fitch03, Achilles04,Rohde07,Perina12,Perina13}. There, an initial state is split uniformly into multiple modes and the presence of photons in each mode is measured by an on-off detector. Such detector delivers only binary information, i.e. a ``click" or ``no-click" \cite{Eisaman11, Lundeen09}. 
In such multiplexing systems, the click statistics however differs from the true photon-number statistics \cite{Sperling12a}. The Mandel $Q_{\rm M}$ parameter ceases to be applicable, requiring a new parameter to properly manifest the nonclassical statistics.
In this regard, Sperling {\it et al.} have proposed the binomial parameter $Q_{\rm B}$ \cite{Sperling12b} defined as
\begin{equation}
Q_{\rm B}=N\frac{\langle (\Delta c)^{2} \rangle}{\langle c \rangle\left(N-\langle c \rangle\right)}-1,
\label{QB}
\end{equation}
where $\langle c \rangle$ and $\langle (\Delta c)^{2} \rangle$ denote the classical mean value and variance of the click statistics, and $N$ is the number of multiplexed modes. They showed that states exhibiting a sub-binomial click statistics, identified by $Q_{\rm B}<0$, are nonclassical. Therefore, measuring $Q_{\rm B}$ can be used to test the nonlcassicality. Despite successful experimental demonstration of the sub-binomial click statistics \cite{Bartley13, Heilmann16} and practical use of the click-statistics for other tasks \cite{Sperling13, Sperling14, Sperling15, Luis15, Lipfert15}, the parameter, however, only applies when the click-counting probabilities over $N$ modes are all equal. To assure this requirement is non-trivial in general. Experimental imperfections, not the statistical fluctuations that keep the average values same, may destroy the required balance, so $``Q_{\rm B}<0"$ is no longer a reliable measure for the nonclassicality.
Moreover, the experimental setup where $Q_{\rm B}$ applies requires at least two or preferably a larger number of on-off detectors. This limits its versatile uses in fully integrated quantum optical circuits. Therefore, novel multiplexing scenarios that require the least experimental resources are highly demanded. In these novel scenarios the constraint on the uniform distribution of the click-counting probabilities can be alleviated, thus requiring the introduction of a new parameter.

\begin{figure}[t]
\centering
\includegraphics[width=8.7cm]{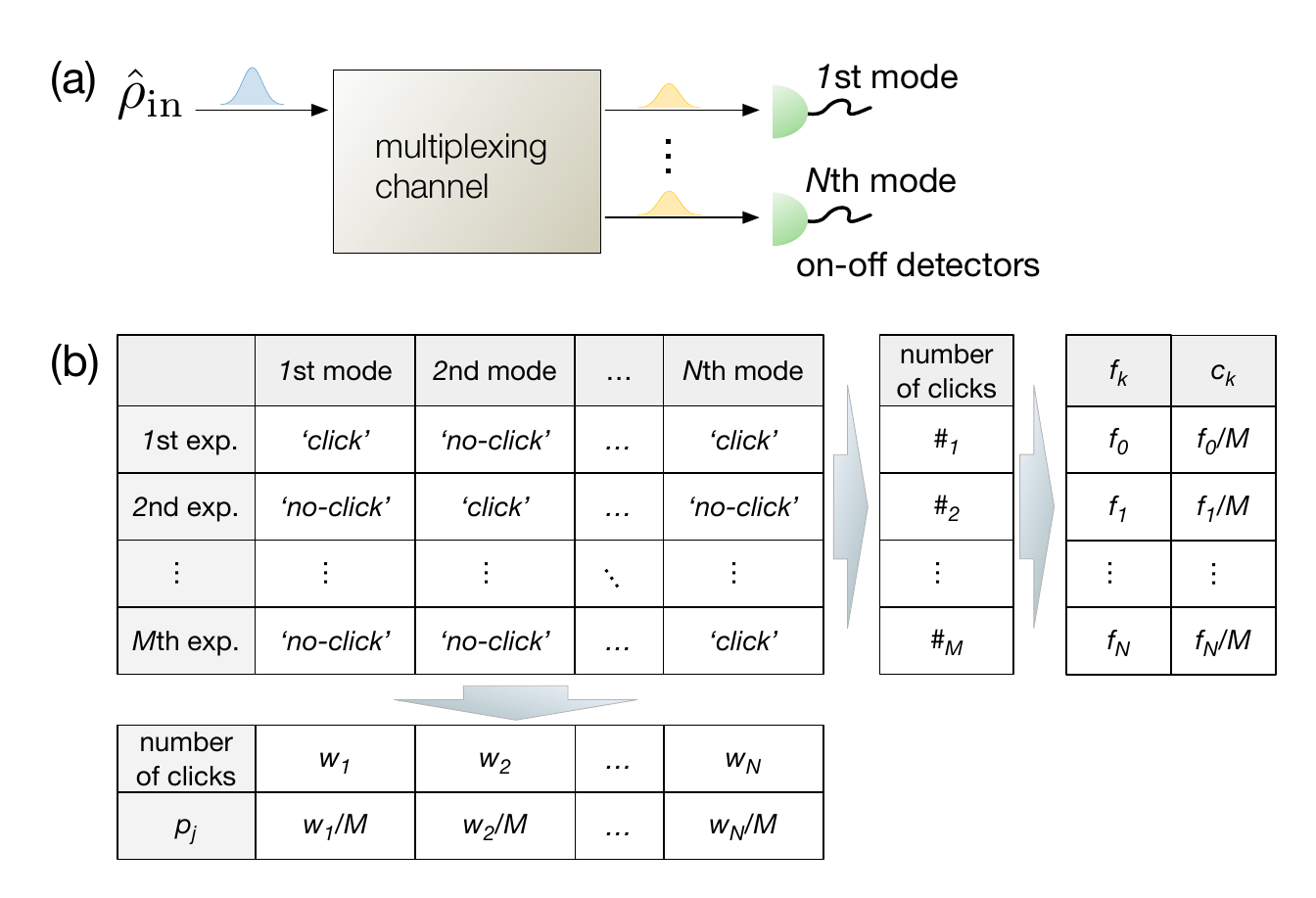}
\caption{
{\bf (a)} A general multiplexing model, where the incoming light is split into $N$ modes in which on-off detections take place.
{\bf (b)} Example of experimental data expected to be obtained from an arbitrary multiplexing detection scheme, where we have repeated the measurement $M$ times and detected either a ``click'' or ``no-click'' in each mode. Each experiment corresponds to a row. Detections over different modes correspond to different columns. For each experiment (or row), we count the number of click-events, and calculate $f_{k}$, the number of experiments that led to $k$ click-events. For each mode (or column), we count the number of click-events, and calculate $w_{j}$, the number of the click events occurred in $j^{\rm th}$ mode. From the distribution $f_{k}$ and $w_{j}$, we generate two kinds of click statistics via $c_{k}=f_{k}/M$ and $p_{j}=w_{j}/M$, respectively.  
}
\label{fig1} 
\end{figure}

In this work we introduce the general parameter $Q_{\rm PB}$ to characterize the click statistics obtained from on-off detectors with a {\it general} multiplexing model where unbalance among $N$ modes is allowed [see Fig.~\ref{fig1}]. It is shown that a measure of $Q_{\rm PB}<0$ unambigously indicates that a state of light is nonclassical. 
$Q_{\rm PB}$ constitutes a more practical nonclassical test scheme in that any unbalances in the detection modes are fully respected, thus enabling the use of arbitrary multiplexing schemes.
We also highlight the significance of our new parameter by examining several states of light in an on-chip platform that employs a ring resonator with only a single on-off detector.
We expect the new parameter will open a new route to explore nonclassicality of light in nanophotonic platforms with minimal experimental restrictions.

\section{Poisson-binomial parameter}
We introduce the Poisson-binomial parameter $Q_{\rm PB}$ defined as
\begin{equation}
Q_{\rm PB}
=N\frac{\langle \left(\Delta c \right)^{2}\rangle}{\langle c \rangle \left(N-\langle c \rangle\right)-N^{2}\sigma^{2}}-1,
\label{QPB}
\end{equation}
where the click statistics $c_{k}$ is characterized by
\begin{align*}
\langle c \rangle =& \sum_{k=0}^{N} k c_{k}, ~~~~~~~~~~~~
\langle \left(\Delta c\right)^{2} \rangle = \sum_{k=0}^{N} \left(k-\langle c\rangle\right)^{2} c_{k},\nonumber
\end{align*}
and, contrary to $Q_{\rm B}$ of Eq.~(\ref{QB}), an additional statistics $p_{j}$ is considered with
\begin{align*}
m=\frac{1}{N}\sum_{j=1}^{N}p_{j}, ~~~~~~
\sigma^{2}=&\frac{1}{N}\sum_{j=1}^{N}\left(p_{j}-m\right)^{2},\nonumber
\end{align*}
where $p_{j}$ denotes a probability of the ``click" event to occur in the $j^{\rm th}$ mode, and $\sigma^{2}$ represents unevenness of $p_{j}$'s compared to their average value $m$. In the denominator of Eq.~(\ref{QPB}), an additional term $N^{2}\sigma^{2}$, absent in $Q_{\rm B}$ of Eq.~(\ref{QB}), respects any unbalanced splitting or different detection probabilities among the $N$ modes.

Before introducing basic properties of $Q_{\rm PB}$, it is worth to discuss how these statistical quantities can be extracted from an experiment. In a general multiplexing detection setup [Fig. \ref{fig1}(a)], the final raw data of the experiment are expected to be as shown in Fig.~\ref{fig1}(b). We suppose that the experiment has been repeated $M$ times using a given input state of light. The click events over $N$ modes are recorded. We then obtain the probabilities $c_{k}$ and $p_{j}$ via $c_{k}=f_{k}/M$ and $p_{j}=w_{j}/M$, where $f_{k}$ is the number of experiments that led to $k$ clicks, and $w_{j}$ is the number of the click events occurred in the $j^{\rm th}$ mode. Note, regardless of the kind of multiplexing scheme, such data table should always be the final outcome from the experiment. Furthermore, the experimental measure of $Q_{\rm PB}$ is only based on the observed statistics, which requires neither {\it a priori} knowledge concerning the system parameters nor loss rates. This substantially reduces the experimental efforts required for analysing systematic uncertainties.


{\it Basic Properties.---}
The parameter $Q_{\rm PB}$ reveals several important properties as follows. 

i) $Q_{\rm PB}$ is zero for light having Poisson-binomial click statistics $c_{k}$, for which $\langle c\rangle=\sum_{j=1}^{N}p_{j}$ and $\langle (\Delta c)^{2}\rangle=\sum_{j=1}^{N}p_{j}(1-p_{j})$ as properties of the Poisson-binomial distribution -- the distribution of the number of successes in a sequence of $N$ independent yes/no experiments with success probabilities $p_{j}$'s \cite{Wang93}. A coherent state is a typical example.

ii) $Q_{\rm PB}$ becomes $Q_{\rm B}$ when $p_{j}=p~\forall j$, immediately resulting in $\sigma^{2}=0$. $Q_{\rm PB}$ also converges to $Q_{\rm M}$ when $N\rightarrow \infty$, in which case $\sigma^{2}\rightarrow0$, $N/(N-\langle c\rangle)\rightarrow1$ and $\langle (\Delta c)^{2} \rangle/\langle c \rangle \rightarrow \langle (\Delta n)^{2} \rangle/\langle n\rangle$.

iii) $Q_{\rm PB}$ is greater than or equal to zero for any classical state of light. Inversely, $Q_{\rm PB}<0$ indicates that the initial state is nonclassical, {\it i.e.}, nonclassicality can be directly observed from the sub-Poisson-binomial behaviour of the click statistics. The proof is given below. 

Following the analysis given in Ref.~\onlinecite{Sperling12a}, the probability $c_{k}$ can be generally written as 
\begin{align*}
c_{k}
=\text{Tr}[\hat{\Pi}_{k}\left(\hat{U}\hat{\rho}_{\rm in}\hat{U}^{\dagger}\right)]
=\int d^{2}\alpha P(\alpha) \bra{\alpha}\hat{U}^{\dagger}\hat{\Pi}_{k}\hat{U}\ket{\alpha},\nonumber
\end{align*}
where $P(\alpha)$ is the Glauber-Sudarshan $P$ representation of an initial state $\hat{\rho}_{\rm in}$ \cite{Scully97}, $\hat{\Pi}_{k}$ denotes the on-off detector operator that triggers $k$ clicks \cite{Rohde07, Sperling12a}, and $\hat{U}$ represents the multiplexing transformation that splits a single-mode initial light into $N$ modes. The on-off detector operator $\hat{\Pi}_{k}$ can be written as
\begin{align*}
\hat{\Pi}_{k}=\sum_{\vert\vec{d}\vert=k}\bigotimes_{\{j\vert d_{j}=1\}} \big(\hat{1} - e^{-(\eta_{j} \hat{n}_{j}+\nu_{j})} \big)\bigotimes_{\{j\vert d_{j}=0\}} e^{-(\eta_{j} \hat{n}_{j}+\nu_{j})},\nonumber
\end{align*}
where $\vec{d}=\{ d_{1},\cdots,d_{N}\}$ is the trigger vector with $d_{j}\in\{0,1\}$ \cite{Rohde07, Sperling12a}, $\eta_{j}=(1-\gamma_{j})\xi_{j}$ denotes the overall quantum efficiency including the individual channel-loss $\gamma_{j}$ and detection efficiency $\xi_{j}$, and $\nu_{j}$ represents the dark count probability for the $j^{\rm th}$ mode. This describes a general multiplexing scenario, where $u_{j}$ represents the distribution probability with $\sum_{j=1}^{N}\abs{u_{j}}^{2}=1$, for example, $\ket{\alpha,0,\cdots,0}\rightarrow \ket{u_{1}\alpha,u_{2}\alpha,\cdots,u_{N}\alpha}$. Therefore, we rewrite
\begin{equation}
c_{k} =\left\langle :  \sum_{\vert\vec{d}\vert=k}\prod_{\{j\vert d_{j}=1\}} \hat{p}_{\rm click}^{(j)} \prod_{\{j\vert d_{j}=0\}} \hat{p}_{\rm no-click}^{(j)}: \right\rangle,
\label{ck}
\end{equation}
where the :$\cdot$: denotes the normal ordering prescription \cite{Scully97, Sperling12a}, $\hat{p}_{\rm click}^{(j)}=\hat{1} - e^{-\left(\eta_{j} \abs{u_{j}}^{2}\hat{n}+\nu_{j}\right)}$, and $\hat{p}_{\rm no-click}^{(j)}=\hat{1}-\hat{p}_{\rm click}^{(j)}$. 
Here, Eq.~(\ref{ck}) represents a quantum version of the Poisson-binomial distribution, and becomes that of the binomial distribution when $N$ modes are uniformly distributed, {\it i.e.}, $u_{j}=u$, $\eta_{j}=\eta$, and $\nu_{j}=\nu~\forall j$. 

After some algebra, it can be shown that 
\begin{align*}
\langle c\rangle =&\bigbraket{: \sum_{j=1}^{N} \hat{p}_{\rm click}^{(j)} :},\nonumber\\
\langle (\Delta c)^{2}\rangle 
=&\bigbraket{:\sum_{j=1}^{N}\hat{p}_{\rm click}^{(j)} \left(\hat{1}-\hat{p}_{\rm click}^{(j)}\right):}+\bigbraket{:\left(\Delta\sum_{j=1}^{N}\hat{p}_{\rm click}^{(j)}\right)^{2}:}, \nonumber
\end{align*}
where $\langle:(\Delta\hat{O})^{2}:\rangle=\langle:\hat{O}^{2}:\rangle-\langle:\hat{O}:\rangle^{2}$ represents the variance of an operator $\hat{O}$, and the relation $\sum_{k=0}^{N}k^{2}c_{k}=\braket{:\sum_{j=1}^{N}\hat{p}_{\rm click}^{(j)} \left(\hat{1}-\hat{p}_{\rm click}^{(j)}\right):}+\braket{:\left(\sum_{j=1}^{N}\hat{p}_{\rm click}^{(j)}\right)^{2}:}$ is used. The probability $p_{j}$, on the other hand, can be obtained by
\begin{align*}
p_{j}={\rm Tr}[\hat{\pi}_{j}\left(\hat{U}\hat{\rho}_{\rm in}\hat{U}^{\dagger}\right)]
=\bigbraket{: \hat{p}_{\rm click}^{(j)} :},
\end{align*}
where $\hat{\pi}_{j}=\hat{1}-e^{-(\eta_{j}\hat{n}_{j}+\nu_{j})}$, so we calculate the classical mean and variance of the distribution $\{ p_{j}\}$ as
\begin{align*}
m=&\frac{1}{N}\bigbraket{: \sum_{j=1}^{N}\hat{p}_{\rm click}^{(j)} :},\nonumber\\
\sigma^{2}=&\frac{1}{N} \sum_{j=1}^{N}\bigbraket{: \hat{p}_{\rm click}^{(j)} :}^{2}-\left( \frac{1}{N}\bigbraket{: \sum_{j=1}^{N}\hat{p}_{\rm click}^{(j)} :}\right)^{2}.\nonumber
\end{align*}
Then, Eq.~(\ref{QPB}) can be explicitly written as 
\begin{equation}
Q_{\rm PB}
=\frac{\bigbraket{:\left(\Delta\sum_{j=1}^{N}\hat{p}_{\rm click}^{(j)}\right)^{2}:}-\sum_{j=1}^{N}\bigbraket{:\left(\Delta \hat{p}_{\rm click}^{(j)}\right)^{2} :}}
{\sum_{j=1}^{N}\bigbraket{: \hat{p}_{\rm click}^{(j)} :}\left(1-\bigbraket{: \hat{p}_{\rm click}^{(j)} :}\right) }.
\label{rQPB}
\end{equation}
Here, the denominator of Eq.~(\ref{rQPB}) is always positive, whereas the numerator can be recast as $\sum_{j\neq k}^{N}\braket{:{\rm Cov}\left(\hat{p}_{\rm click}^{(j)},\hat{p}_{\rm click}^{(k)}\right):}$. For classical states for which the $P(\alpha)$ distribution exhibits the properties of a classical probability distribution, the covariance $\braket{:{\rm Cov}\left(\hat{p}_{\rm click}^{(j)},\hat{p}_{\rm click}^{(k)}\right):}=\braket{:\hat{p}_{\rm click}^{(j)}\hat{p}_{\rm click}^{(k)}:}-\braket{:\hat{p}_{\rm click}^{(j)}:}\braket{:\hat{p}_{\rm click}^{(k)}:}$ is non-negative since $\hat{p}_{\rm click}^{(j)}$ and $\hat{p}_{\rm click}^{(k)}$ behave in the same direction with respect to $\alpha$, regardless of imperfections represented by $\eta_{j}$ and $\nu_{j}$. Thus, for any classical state, $Q_{\rm PB}\ge0$. Such inequality, imposed by the assumption that the field is classical, provides the sufficient condition by its violation for states to be said `nonclassical'. The $Q_{\rm B}$ of Eq.~(8) in Ref.~\onlinecite{Sperling12b} is reproduced when $\hat{p}_{\rm click}^{(j)}=\hat{p}_{\rm click}~\forall j$, in which case the numerator and denominator of Eq.~(\ref{rQPB}) are recast as 
$N(N-1)\braket{:\left(\Delta \hat{p}_{\rm click}\right)^{2}:}$, and $N\braket{:\hat{p}_{\rm click}:}\left(1-\braket{:\hat{p}_{\rm click}:}\right) = \braket{c}\left(1-\braket{c}/N\right)$, respectively, thus resulting in $Q_{\rm PB}=Q_{\rm B}$.

\begin{figure}[b]
\centering
\includegraphics[width=9cm]{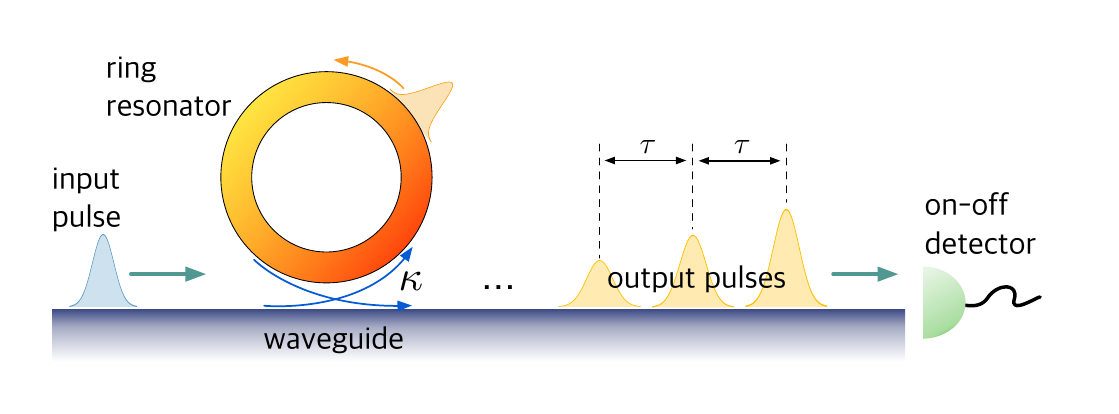}
\caption{A ring resonator coupled to waveguide setup for the time-bin multiplexing detection scheme.
Via the coupling strength $\kappa$ that depends on the geometric distance and mode-matching conditions, the propagating mode couples into the ring resonator and the trapped field escapes to the waveguide. The output pulses are separated in time by $\tau$, the round-trip delay, which depends on the ring size and speed of the propagating mode. 
}
\label{fig2} 
\end{figure}

As in the binomial parameter $Q_{\rm B}$ \cite{Sperling12b}, the nonclassical test with the Poisson-binomial parameter $Q_{\rm PB}$ works even when $N=2$, providing the simplest nonclassicality test scheme. The number of experiments $M$, on the other hand, needs to be large enough so that the probabilities $c_{k}$ and $p_{j}$ are more reliable as in usual experiments where statistical quantities are examined.

\section{Optical ring resonator setup}
A typical example for which our new parameter $Q_{\rm PB}$ is suitable is a fiber-loop detector setup \cite{Banaszek03, Rehacek03} where unbalances are naturally introduced due to the asymmetric mode-division and losses. 
As a similar scheme, but towards on-chip quantum devices at the nanoscale, an optical ring resonator setup is considered in this work to demonstrate the feasibility of the parameter $Q_{\rm PB}$. Such resonators can be implemented in nanophotonic circuits \cite{Pernice12}. As illustrated in Fig.~\ref{fig2}, a waveguide supporting a propagating mode is coupled to a ring resonator via an evanescent field, and a single on-off detector is placed at the end of the waveguide. Herein, the incoming light in the waveguide may enter a ring resonator, and then starts multiple round-trips until it escapes back to the waveguide. This setup leads to the emergence of consecutive output pulses along the waveguide provided that the initial pulse width is much narrower than the round-trip time $\tau$. Each pulse corresponds to a single multiplexed mode. The outgoing pulses have different amplitudes that depend on the coupling strength $\kappa$ and the losses induced by different travel lengths, thus resulting in different probabilities of reaching the detector. The coupling strength $\kappa$ determines the fraction of amplitudes for each pulse, {\it i.e.}, 
\begin{align*}
\abs{u_{1}}^{2}=1-\kappa,~ {\rm and}~
\abs{u_{j}}^{2}=\kappa^{2}\left(1-\kappa\right)^{j-2}~{\rm for}~j\ge2\nonumber
\end{align*}
with $\sum_{j=1}^{N=\infty}\abs{u_{j}}^{2}=1$, whereby an infinite number of modes is produced in principle. For practical relevance, only the first $N_{\rm trc}$ output pulses are to be selectively detected, yet the basic three properties of $Q_{\rm PB}$ still hold with the truncation $N_{\rm trc}$ (see Appendix).
Individual quantum efficiency $\eta_{j}$ is determined by the overall loss rate $\gamma_{j}$ due to scattering and intrinsic loss, and also by the detection efficiency $\xi$ of a single detector. For simplicity, but without loss of generality for the purpose of this work, we assume that the overall efficiencies are all equal and the dark count rate is negligible, {\it i.e.}, $\eta_{j}=\eta~\forall j$ and $\nu\approx 0$, so that the unbalance is induced by different amplitudes of the output pulses. This setup serves as a multiplexing scheme when the duration $\tau$ of a single round trip in a resonator is longer than the dead time of detector \cite{Pernice12}. In this setup, the balance among the output pulses is naturally broken, so only our new parameter $Q_{\rm PB}$ offers an adequate measure for nonclassicality.

\begin{figure}[b]
\centering
\includegraphics[width=8cm]{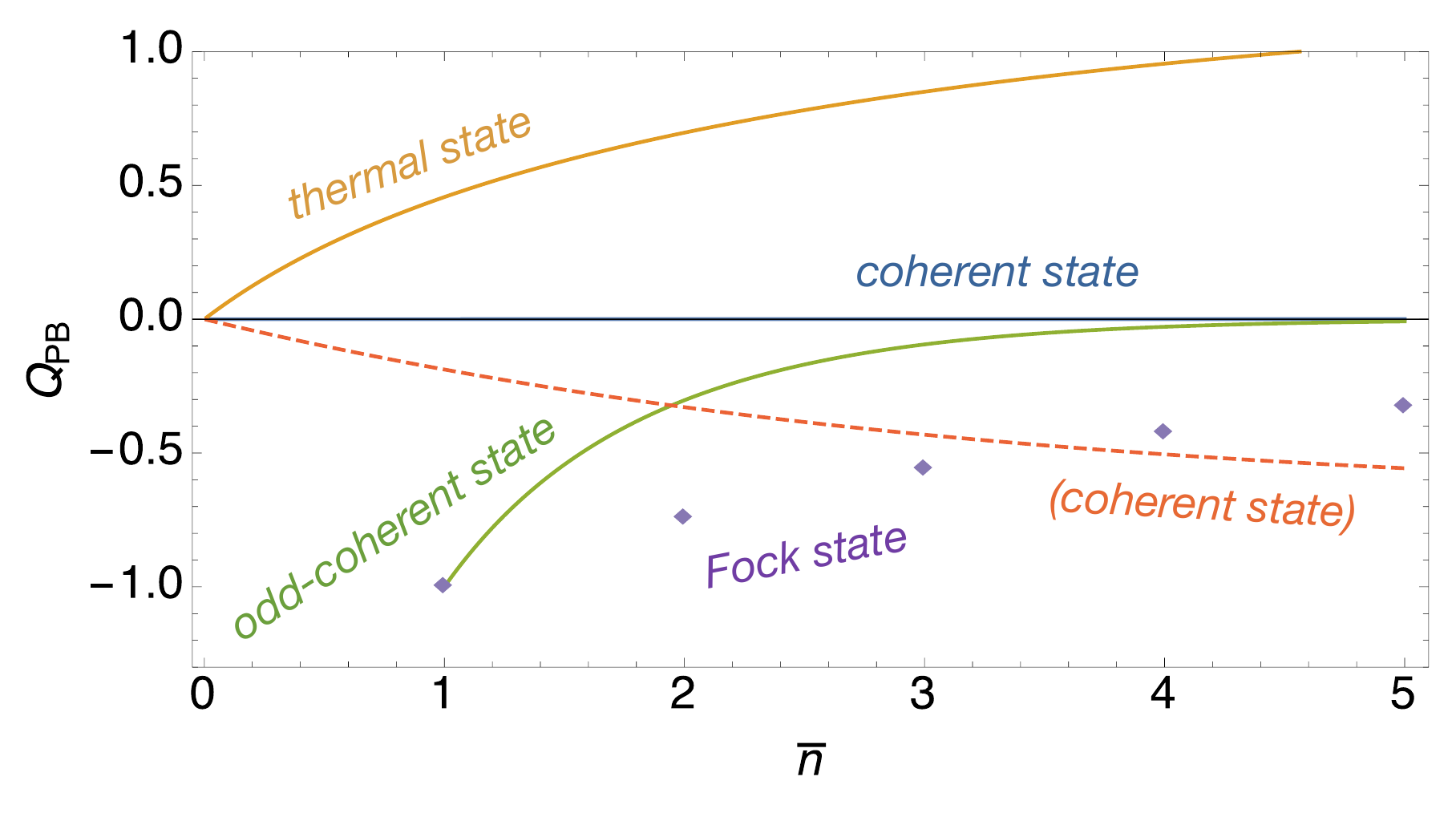}
\caption{$Q_{\rm PB}$ for a classical coherent, thermal, Fock, and odd-coherent states in the ring resonator setup. The positive (negative) $Q_{\rm PB}$ exhibits the super (sub)-Poisson-binomial click statistics. The dashed line represents $Q_{\rm B}$ for the coherent state in the same setup.
}
\label{fig3} 
\end{figure}

Now we investigate the behaviour of $Q_{\rm PB}$ for well-known classical and quantum input states in the ring resonator setup. The example classical states include a coherent state $\ket{\alpha}$ and thermal state $\hat{\rho}_{\rm th}$, which have the Poisson, and Bose-Einstein statistics for the photon number distribution, respectively, {\it i.e.}, $\rho_{\rm coh}(n)=e^{-\bar{n}}\bar{n}^{n}/n!$ and $\rho_{\rm th}(n)=\bar{n}^{n}/\left(1+\bar{n}\right)^{n+1}$, where $\bar{n}$ is the mean photon number. The example quantum states, on the other hand, include a Fock state $\ket{m}$ and odd-coherent state $\left(\ket{\alpha}-\ket{-\alpha}\right)/N_{-}^{1/2}$ whose statistics are given as $\rho_{\rm Fock}(n)=\delta_{n m}$ and $\rho_{\rm oc}(n)=4 e^{-\abs{\alpha}^{2}}\abs{\alpha}^{2n} \left(1-(-1)^{n}\right)/2n! N_{-}$, respectively, where $N_{-}=2(1-e^{-2\abs{\alpha}^{2}})$.

To see a response of the ring resonator setup to the input states, we perform Monte Carlo simulations to mimic the probabilistic multiplexing and detection mechanisms, whereby the conditional probabilities $C(k\vert n)$ and $P(j\vert n)$ are obtained for $0\le k \le N_{\rm trc}$, $1\le j\le N_{\rm trc}$, and $0\le n\le n_{\rm max}$. 
Here, the conditional probability $C(k\vert n)$ denotes the probability of delivering $k$ clicks over the truncated $N_{\rm trc}$ modes for an incident $n$-photon, whereas $P(j\vert n)$ is the probability of the click to be triggered in the $j^{\rm th}$ mode. Incorporating the latters with the input photon-number distribution $\rho_{\rm in}(n)$, we obtain the click statistics via
\begin{align*}
c_{k}=\sum_{n=0}^{n_{\rm max}}C(k\vert n)\rho_{\rm in}(n),~\text{and}~
p_{j}=\sum_{n=0}^{n_{\rm max}}P(j\vert n)\rho_{\rm in}(n).
\end{align*}
Here $n_{\rm max}=30$ is chosen such that the higher-photon-number contribution of an initial state, addressed by $\rho_{\rm in}(n>n_{\rm max})$, is negligible with respect to the mean photon number of the example states.

\begin{figure}[t]
\centering
\includegraphics[width=8cm]{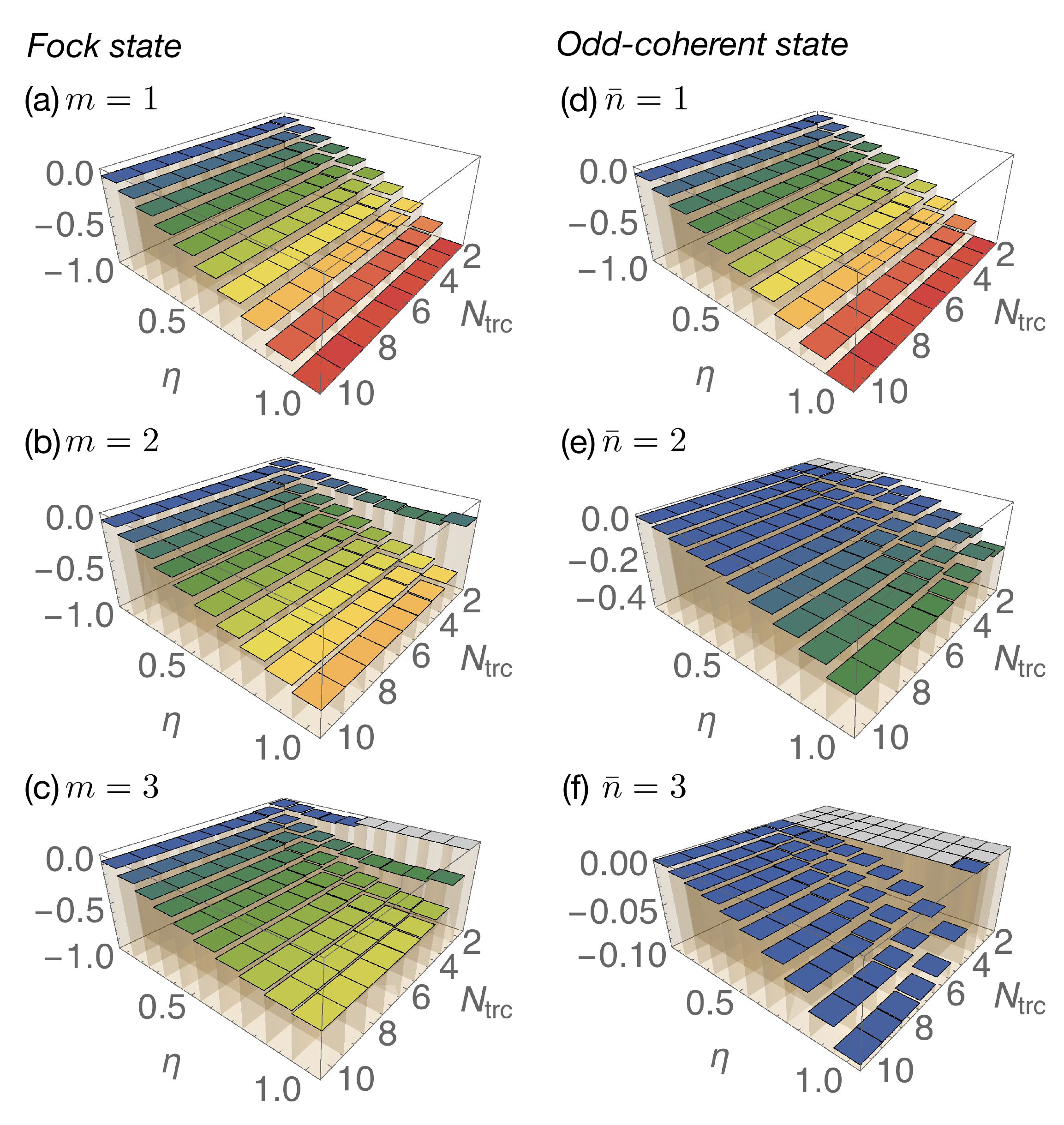}
\caption{The effect of efficiency $\eta$ and the number $N_{\rm trc}$ of pulses are investigated for the Fock states with $m=1,2,3$ (left column) and the odd-coherent states with $\bar{n}=1,2,3$ (right column). Here $\kappa=0.6$.
}
\label{fig4} 
\end{figure}

In Fig.~\ref{fig3}, we present the behaviours of $Q_{\rm PB}$ with the mean photon number $\bar{n}$ of the input states from the observed click statistics based on $M=10^6$ Monte Carlo simulations, where $\kappa=0.6$, $\eta=1$, and $N_{\rm trc}=10$ are chosen. The $Q_{\rm PB}$ measures for the coherent state the boundary of the nonclassical test in the considered setup. Here, a small deviation from zero, $Q_{\rm PB}\approx0$, arises due to the finite number $M$ of measurements performed, so it will become zero when $M\rightarrow\infty$. It is clearly shown that the thermal state exhibits super-Poisson-binomial click statistics, whereas the quantum states manifest in sub-Poisson-binomial click statistics, regardless of the mean photon number. Thus, the Poisson-binomial parameter $Q_{\rm PB}$ constitutes a more reliable nonclassicality test using the click statistics in the considered on-chip multiplexing detection setup, as compared to the binomial parameter $Q_{\rm B}$ which shows negative values even for the coherent state input due to the unbalance involved (see the dashed line in Fig.~\ref{fig3}). 

\begin{figure}[t]
\centering
\includegraphics[width=7.3cm]{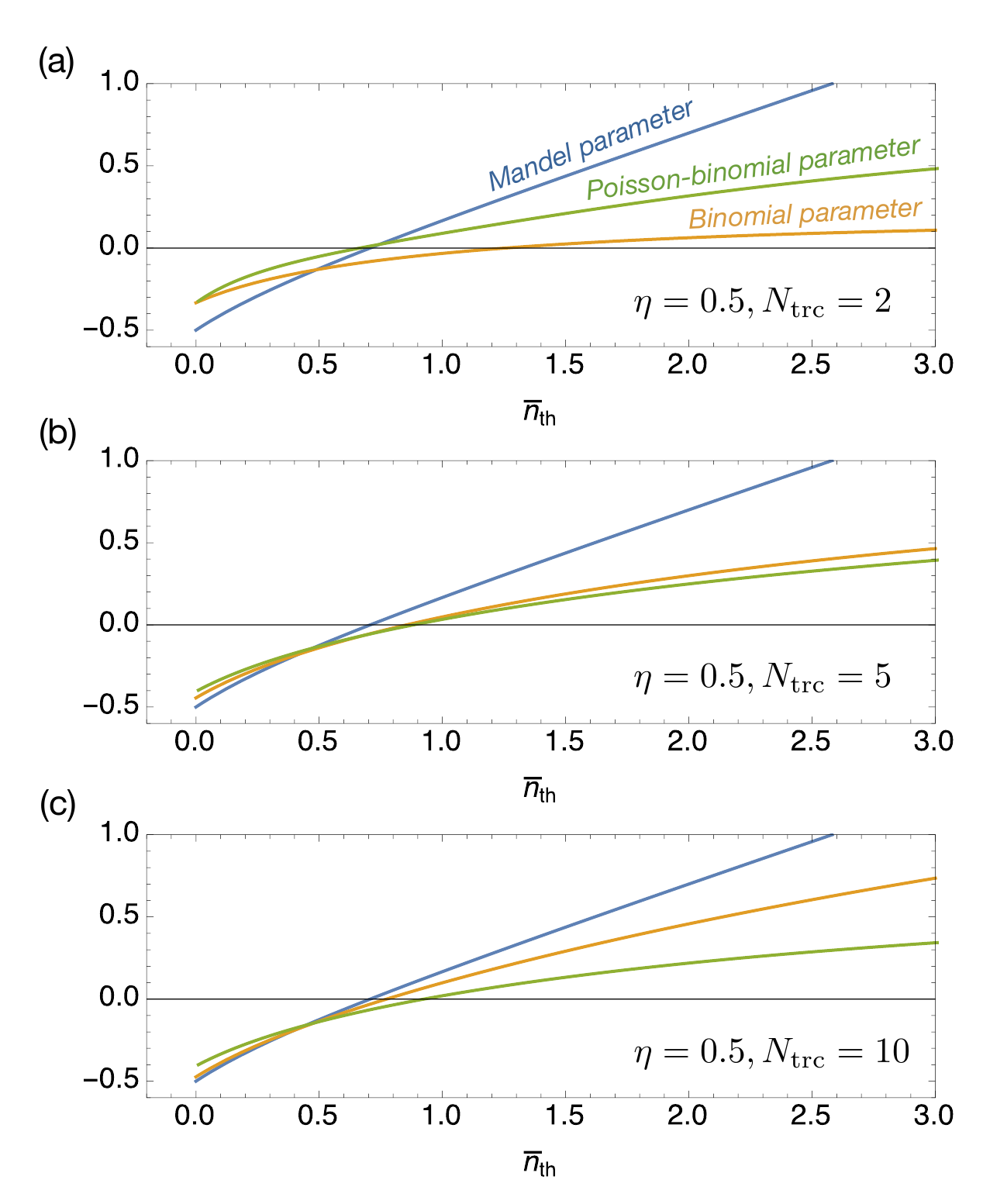}
\caption{
For SPATSs, $Q_{\rm PB}$ is calculated by the click statistics obtained from Monte Carlo simulations where we choose $\kappa=0.6$, and $n_{\rm max}=30$, whereas $Q_{\rm M}$ and $Q_{\rm B}$ are calculated via the closed expressions provided by Ref.~\onlinecite{Sperling12b}.
}
\label{fig5} 
\end{figure}

Looking more closely, one can see that the thermal state has a monotonic trend in $Q_{\rm PB}$ with the mean photon number $\bar{n}$, and also the $Q_{\rm PB}$ for the odd-coherent state approaches the coherent state case. The same behaviours are also observed in both $Q_{\rm M}$ and $Q_{\rm B}$, implying that the observed statistics strongly reflects the characteristics of the initial photon number distribution with $\bar{n}$. The Fock state case on the other hand increases with the photon number, but this arises due to the finite number of $N_{\rm trc}$ and $n_{\rm max}$ used in Monte Carlo simulations. Namely it will be closer to $-1$ as both $N_{\rm trc}$ and $n_{\rm max}$ increase further, where $Q_{\rm PB}\rightarrow Q_{\rm M}$. 

In Fig.~\ref{fig4}, we investigate the effect of the overall efficiency $\eta$ and the number $N_{\rm trc}$ of output pulses to be measured for the example quantum input states. It is interesting to see that both states exhibit sub-Poisson-binomial click statistics regardless of $\eta$ and $N_{\rm trc}$ when the mean photon number equals one [see Figs.~\ref{fig4}(a) and (d)]. As the input energy increases, on the other hand, higher $N_{\rm trc}$ is required to successfully capture the nonclassical features of the click statistics [see Figs.~\ref{fig4}(b),(c),(e), and (f)]. There is a remarkable observation worth stressing; even when the overall efficiency $\eta$ is significantly smaller than unity, the sub-Poisson-binomial statistics still survives at the cost of measuring more output pulses. Therefore, the measure of $Q_{\rm PB}$ is practical in that it can be used even when significant losses are inevitably involved.

In addition, one may be interested in applying the nonclassicality test via $Q_{\rm PB}$ to other quantum states such as a single-photon-added thermal state (SPATS) whose photon number distribution is given as $\rho_{\rm SPATS}(n)={\rm Tr}[\hat{n}\hat{\rho}_{\rm SPATS}]=\frac{n}{\bar{n}_{\rm th}(\bar{n}_{\rm th}+1)} \left(\frac{\bar{n}_{\rm th}}{\bar{n}_{\rm th}+1}\right)^{n}$, where $\bar{n}_{\rm th}$ denotes the mean thermal photon number, while the total mean photon number is given as $\bar{n}=1+2\bar{n}_{\rm th}$ \cite{Agarwal92}. For the latter, there has been an interesting observation that at a certain range of $\bar{n}_{\rm th}$, the binomial parameter $Q_{\rm B}$ captures the nonclassicality of SPATSs that the $Q_{\rm M}$ cannot identify even with the photon-number-resolving capability \cite{Sperling12b}. Somewhat surprisingly, we also find that our new parameter extends the range where one can identify the nonclassical feature of light. Such behaviours depend on various parameters such as $N_{\rm trc}$ and $\eta$ as shown in Fig.~\ref{fig5}, where we present $Q_{\rm PB}$ obtained from Monte Carlo simulations for the ring-resonator setup as compared with $Q_{\rm M}$ and $Q_{\rm B}$ whose closed expressions are respectively given as
$Q_{\rm M}= \eta (\bar{n}_{\rm th}^{2}-\frac{1}{2})/(\bar{n}_{\rm th}+\frac{1}{2})$, and $Q_{\rm B}= (N_{\rm trc}-1)\left(I\left(\frac{2\eta}{N_{\rm trc}}\right)-I\left(\frac{\eta}{N_{\rm trc}}\right)^{2}\right)/\left(I\left(\frac{\eta}{N_{\rm trc}}\right)[1-I\left(\frac{\eta}{N_{\rm trc}}\right)]\right)$, where $I\big( \lambda \big)=(1-\lambda)/(1+\lambda \bar{n}_{\rm th})^{2}$. These expressions are taken from Ref.~\onlinecite{Sperling12b}. From Figs.~\ref{fig5}(a)-(c), one can see that $Q_{\rm PB}$ captures the nonclassical features at a certain range of $\bar{n}_{\rm th}$, in which both $Q_{\rm M}$ and $Q_{\rm B}$ have positive values. This is observed when $N_{\rm trc}>5$ with respect to the parameters we used. 

\section{Remarks}
In this work, we have introduced the general parameter $Q_{\rm PB}$ to measure the nonclassicality of an input state of light in a general multiplexing detection scheme. This scheme considerably reduces experimental restrictions as compared to the Mandel $Q_{\rm M}$ and binomial $Q_{\rm B}$ parameters. 
It is worth mentioning that our method does not require the reconstruction of the true photon number statistics that is directly based on {\it a priori} knowledge of system parameters and loss rates. Furthermore, the optimization of the multiplexing scheme can be also relaxed, so that our new parameter would pave the wave for more practical quantum measurement for various quantum information applications such as quantum cryptography and quantum metrology.
Further understanding of the click statistics provided in this work will stimulate a variety of future studies on the measurement of quantum correlations, quadrature phases, or multimode properties, associated with other interesting quantum states as well as imperfections (e.g., the dark count and the sampling error) not explicitly considered in this work.
Developing the parameter to quantify the nonlcassicality of light and also to distinguish quantum states showing the super-Poisson-binomial click statistics (e.g., a squeezed vacuum state) is also an interesting future work.

\section*{ACKNOWLEDGMENTS}
C. Lee thanks Su-Yong Lee for comments. 

\section*{APPENDIX}
The ring-resonator setup produces an infinite number of output pulses, but only the first $N_{\rm trc}$ output pulses are to be detected for practical relevance. Here, we show that the basic properties of $Q_{\rm PB}$ still hold with the truncation $N_{\rm trc}$, and also examine the case of $N_{\rm trc}=1$. To this end, we replace $N$ by $N_{\rm trc}$ in Eq.~(3), and then it reads as
\begin{align*}
Q_{\rm PB}=N_{\rm trc}\frac{\langle \left(\Delta c \right)^{2}\rangle}{\langle c \rangle \left(N_{\rm trc}-\langle c \rangle\right)-N_{\rm trc}^{2}\sigma^{2}}-1,
\end{align*}
where
$\langle c \rangle = \sum_{k=0}^{N_{\rm trc}} k c_{k}$,
$\langle \left(\Delta c\right)^{2} \rangle = \sum_{k=0}^{N_{\rm trc}} \left(k-\langle c\rangle\right)^{2} c_{k}$,
$m=\frac{1}{N_{\rm trc}}\sum_{j=1}^{N_{\rm trc}}p_{j}$, and
$\sigma^{2}=\frac{1}{N_{\rm trc}}\sum_{j=1}^{N_{\rm trc}}(p_{j}-m)^{2}$.
Below we prove that the basic three properties of $Q_{\rm PB}$ are still preserved with the truncation.
\begin{enumerate}
\item For light having Poisson-binomial click statistics $c_{k}$, 
$\langle c \rangle = \sum_{j=1}^{N_{\rm trc}}p_{j}$ and
$\langle \left(\Delta c\right)^{2} \rangle = \sum_{j=1}^{N_{\rm trc}}p_{j} (1-p_{j})$,
so that $Q_{\rm PB}=0$.\hfill (Q.E.D.)

\item When $p_{j}=p~ \forall j$, $\sigma^{2}=0$, so that
$Q_{\rm PB}=Q_{\rm B}$.\hfill (Q.E.D.)\\
In the limit of $N_{\rm trc}\rightarrow \infty$, it is shown that
$m \rightarrow 0$,
$\sigma^{2} \rightarrow 0$,
$N_{\rm trc}/\left(N_{\rm trc}-\langle c \rangle\right) \rightarrow 1$, and
$\langle \left(\Delta c \right)^{2}\rangle/\langle c \rangle \rightarrow \langle \left(\Delta n \right)^{2}\rangle/\langle n \rangle$,
so that $Q_{\rm PB}\rightarrow Q_{\rm M}$. \hfill (Q.E.D.)

\item With the truncation to $N_{\rm trc}$, Equation (5) remains unaltered except $N\rightarrow N_{\rm trc}$. Thus, the same property with respect to the covariance still holds for classical states for which the $P(\alpha)$ distribution exhibits the properties of a classical probability distribution. That is, it follows for classical states, $Q_{\rm PB}\ge0$ with the truncation of the number of modes. \hfill (Q.E.D.)
\end{enumerate}
Now we examine the behaviour of $Q_{\rm PB}$ when $N_{\rm trc}= 1<N$. In this case, the $Q_{\rm PB}$ can be written as $Q_{\rm PB}=c_{1}(c_{0}+c_{1}-1 )/(1- c_{1})=0$, where $c_{1}=1-c_{0}$, irrespective of input states and the value $p_{1}$. That is, any state, regardless of whether it is quantum or classical, has a Poisson-binomial click statistics, (precisely speaking, it is a binomial statistics because $\sigma^{2}=0$), when a single-mode detection ($N_{\rm trc}=1)$ is used. 



\begin{thebibliography}{99}
\bibitem{Bouwmeester97} D. Bouwmeester, J.-W. Pan, K. Mattle, M. Eibl, H. Weinfurter, and A. Zeilinger, ``Experimental quantum teleportation", Nature {\bf 390}, 575 (1997).
\bibitem{Gisin07} N. Gisin and R. Thew, ``Quantum communication", Nat. Photon. {\bf 1}, 165 (2007).
\bibitem{Ladd10} T. D. Ladd, F. Jelezko, R. Laflamme, Y. Nakamura, C. Monroe, and J. L. O'Brien, ``Quantum computers", Nature {\bf 464}, 45 (2010). 

\bibitem{Dodonov00} V. V. Dodonov, O. V. Man'ko, V. I. Man'ko, and A. W\"unsche, ``Hilbert-Schmidt distance and non-classicality of states in quantum optics", J. Mod. Opt. {\bf 47}, 633 (2000).
\bibitem{Richter02} Th. Richter and W. Vogel, ``Nonclassicality of Quantum States: A Hierarchy of Observable Conditions", Phys. Rev. Lett. {\bf 89}, 283601 (2002).
\bibitem{Kenfack04} A. Kenfack and K. \.{Z}yczkowski, ``Negativity of the Wigner function as an indicator of non-classicality", J. Opt. B: Quantum Semiclass. Opt. {\bf 6}, 396 (2004).
\bibitem{Asboth05} J. K. Asb\'oth, J. Calsamiglia, and H. Ritsch, ``Computable Measure of Nonclassicality for Light", Phys. Rev. Lett. {\bf 94}, 173602 (2005).

\bibitem{Mandel79} L. Mandel, ``Sub-Poissonian photon statistics in resonance fluorescence", Opt. Lett. {\bf 4}, 205 (1979).

\bibitem{Gerrits10} T. Gerrits, S. Glancy, T. S. Clement, B. Calkins, A. E. Lita, A. J. Miller, A. L. Migdall, S. W. Nam, R. P. Mirin, and E. Knill, ``Generation of optical coherent-state superpositions by number-resolved photon subtraction from the squeezed vacuum", Phys. Rev. A {\bf 82}, 031802 (2010).
\bibitem{Gerrits11} T. Gerrits, N. Thomas-Peter, J. C. Gates, A. E. Lita, B. J. Metcalf, B. Calkins, N. A. Tomlin, A. E. Fox, A. L. Linares, J. B. Spring, N. K. Langford, R. P. Mirin, P. G. R. Smith, I. A. Walmsley, and S. W. Nam, ``On-chip, photon-number-resolving, telecommunication-band detectors for scalable photonic information processing", Phys. Rev. A {\bf 84}, 060301 (2011).
\bibitem{Brida12} G. Brida, L. Ciavarella, I. P. Degiovanni, M. Genovese, L. Lolli, M. G. Mingolla, F. Piacentini, M. Rajteri, E. Taralli, and M. G. A. Paris, ``Quantum characterization of superconducting photon counters", New J. Phys. {\bf 14}, 085001 (2012).
\bibitem{Namekata10} N. Namekata, Y. Takahashi, G. Fujii, D. Fukuda, S. Kurimura, and S. Inoue, ``Non-Gaussian operation based on photon subtraction using a photon-number-resolving detector at a telecommunications wavelength", Nat. Photon. {\bf 4}, 655 (2010).

\bibitem{Banaszek03} K. Banaszek and I. A. Walmsley, ``Photon counting with a loop detector", Opt. Lett. {\bf 28}, 52 (2003).
\bibitem{Rehacek03} J. {\v R}eh\'a{\v c}ek, Z. Hradil, O. Haderka, J. Pe{\v r}ina Jr., and M. Hamar, ``Multiple-photon resolving fiber-loop detector", Phys. Rev. A {\bf 67}, 061801 (R) (2003).
\bibitem{Achilles03} D. Achilles, Ch. Silberhorn, C. \'Sliwa, K. Banaszek, and I. A. Walmsley, ``Fiber-assisted detection with photon number resolution", Opt. Lett. {\bf 28}, 2387 (2003). 
\bibitem{Fitch03} M. J. Fitch, B. C. Jacobs, T. B. Pittman, and J. D. Franson, ``Photon-number resolution using time-multiplexed single-photon detectors", Phys. Rev. A {\bf 68}, 043814 (2003).
\bibitem{Achilles04} D. Achilles, Ch. Silberhorn, C. \'Sliwa, K. Banaszek, I. A. Walmsley, M. J. Fitch, B. C. Jacobs, T. B. Pittman, and J. D. Franson, ``Photon-number-resolving detection using time-multiplexing", J. Mod. Opt. {\bf 51}, 1499 (2004).
\bibitem{Rohde07} P. P. Rohde, J. G. Webb, E. H. Huntington, and T. C. Ralph, ``Photon number projection using non-number-resolving detectors", New J. Phys. {\bf 9}, 233 (2007).

\bibitem{Perina12} J. Pe{\v r}ina, Jr., M. Hamar, V. Mich\'alek, and O. Haderka, ``Photon-number distributions of twin beams generated in spontaneous parametric down-conversion and measured by an intensified CCD camera", Phys. Rev. A {\bf 85}, 023816 (2012).
\bibitem{Perina13} J. Pe{\v r}ina, Jr., O. Haderka, V. Mich\'alek, and M. Hamar, ``State reconstruction of a multimode twin beam using photodetection", Phys. Rev. A {\bf 87}, 022108 (2013).


\bibitem{Eisaman11} M. D. Eisaman, J. Fan, A. Migdall, and S. V. Polyakov, ``Invited Review Article: Single-photon sources and detectors", Rev. Sci. Instrum. {\bf 82}, 071101 (2011). 
\bibitem{Lundeen09} J. S. Lundeen, A. Feito, H. Coldenstrodt-Ronge, K. L. Pregnell, Ch. Silberhorn, T. C. Ralph, J. Eisert, M. B. Plenio, and I. A. Walmsley, ``Tomography of quantum detectors", Nat. Phys. {\bf 5}, 27 (2009).

\bibitem{Sperling12a} J. Sperling, W. Vogel, and G. S. Agarwal, ``True photocounting statistics of multiple on-off detectors", Phys. Rev. A {\bf 85}, 023820 (2012).
\bibitem{Sperling12b} J. Sperling, W. Vogel, and G. S. Agarwal, ``Sub-Binomial Light", Phys. Rev. Lett. {\bf 109}, 093601 (2012).

\bibitem{Bartley13} T. J. Bartley, G. Donati, X.-M. Jin, A. Datta, M. Barbieri, and I. A. Walmsley, ``Direct Observation of Sub-Binomial Light", Phys. Rev. Lett. {\bf 110}, 173602 (2013).
\bibitem{Heilmann16} R. Heilmann, J. Sperling, A. Perez-Leija, M. Gr\"afe, M. Heinrich, S. Nolte, W. Vogel, and A. Szameit, ``Harnessing click detectors for the genuine characterization of light states", Sci. Rep. {\bf 6}, 19489, (2016). 

\bibitem{Sperling13} J. Sperling, W. Vogel, and G. S. Agarwal, ``Correlation measurements with on-off detectors", Phys. Rev. A {\bf 88}, 043821 (2013).
\bibitem{Sperling14} J. Sperling, W. Vogel, and G. S. Agarwal, Phys. ``Quantum state engineering by click counting", Rev. A {\bf 89}, 043829 (2014).
\bibitem{Sperling15} J. Sperling, M. Bohmann, W. Vogel, G. Harder, B. Brecht, V. Ansari, and C. Silberhorn, ``Uncovering Quantum Correlations with Time-Multiplexed Click Detection", Phys. Rev. Lett. {\bf 115}, 023601 (2015).
\bibitem{Luis15} A. Luis, J. Sperling, and W. Vogel, ``Nonclassicality Phase-Space Functions: More Insight with Fewer Detectors", Phys. Rev. Lett. {\bf 114}, 103602 (2015).
\bibitem{Lipfert15} T. Lipfert, J. Sperling, and W. Vogel, ``Homodyne detection with on-off detector systems", Phys. Rev. A {\bf 92}, 053835 (2015).

\bibitem{Wang93} Y. H. Wang, "On the number of successes in independent trials", Statistical Sinica, {\bf 3}, 295 (1993).

\bibitem{Scully97} M. O. Scully and M. S. Zubairy, {\it Quantum Optics} (Cambridge University Press, Cambridge, England, 1997).

\bibitem{Pernice12} W. H. P. Pernice, C. Schuck, O. Minaeva, M. Li, G. N. Goltsman, A. V. Sergienko, and H. X. Tang, ``High-speed and high-efficiency travelling wave single-photon detectors embedded in nanophotonic circuits", Nat. Commun. {\bf 3}, 1325 (2012). 

\bibitem{Agarwal92} G. S. Agarwal and K. Tara, ``Nonclassical character of states exhibiting no squeezing or sub-Poissonian statistics", Phys. Rev. A {\bf 46}, 485 (1992).

%
%
%
%
%
%
%
\end{thebibliography}
\end{document}